\documentstyle[12pt]{article}

\begin{document}
\setcounter{page}{1}
\pagestyle{plain} \vspace{1cm}
\begin{center}
\Large{\bf   Phase transition in multi-scalar-singlet extensions of the Standard Model}\\
\small \vspace{1cm} {\bf A.
Tofighi}\footnote{A.Tofighi@umz.ac.ir},\qquad {\bf O. N.
Ghodsi}\qquad {\bf and}
\qquad{\bf M. Saeedhoseini}\\\

\vspace{0.5cm} {\it Department of Physics, Faculty of Basic Sciences,\\
University of Mazandaran,
P. O. Box 47416-95447, Babolsar, Iran}\\

\end{center}
\vspace{1.5cm}
\begin{abstract}
We propose a generalization of the Standard Model $(SM)$ by adding
two real gauge-singlets $S_1, S_2$. The field $S_1$ will improve the
strength of the electroweak phase transition $(EWPT)$. Imposing a 
$Z_2$ symmetry on the field $S_2$ makes this field a possible
candidate for dark matter. Both singlets interact with other
observable fields through Higgs boson. They are allowed to interact
with each other as well. We find that by introducing two different
scalar fields, the model is less vulnerable
to experimental constraints.\\
In this paper, we consider the effects of a heavy scalar$(M_1>M_H)$
on the electroweak phase transition.
 And we present configurations that produce a strong first order
$EWPT$.\\

\end{abstract}
\vspace{1.5cm}

\section{Introduction}
A solution to explain the baryon asymmetry of our universe $[1]$ is based on
 violation of baryon number, $C$ and $CP$ violation as well as a departure from thermal equilibrium.
In a viable model of electroweak baryogenesis
the
departure from thermal equilibrium is realized via a strong first order phase transition $[2]$.
But the $SM$ of particle physics can not provide a strong first order phase transition $[3]$. In addition
the $SM$ does not have a candidate for dark matter ($DM$) as well.\\
Moreover, from the three-loop $\beta$ function for the Higgs self coupling
it is found that
 Higgs vacuum is no longer
stable beyond the scale $ 10^{10}$ $GeV$. Hence we expect some new physics
to appear before this scale $[4]$.\\
Therefore, some new models are required to address these issues.
A popular model is to couple a singlet scalar to Higgs boson $[5-16]$.\\
In Ref. $[17]$ a scheme for classifying models of the electroweak phase transition
has been presented.
One may associate the formation of gravitational waves to a strong first-order phase transition $[18-21]$.\\
But models with an addition of one real singlet can not address all of the shortcomings of
the $SM$.
An interesting class of models are the multi-singlet extensions of the $SM$ models $[22-28]$.
In these models there are more opportunities to satisfy the constraints imposed by experimental findings $[29]$.\\
 In order to study the dynamics of the electroweak
phase transition $EWPT$ one has to resort to techniques from the domain of
thermal field theory $[30-33]$. An essential element is finite-temperature effective potential, which
is a measure of the free energy density of the system. Generally
a loop-level analysis,
 in conjunction with a vigourous Monte-Carlo scan of the
parameter space is needed to unravel the structure of $EWPT$.
However in this work we study the dynamics of the $EWPT$ at the tree level.\\
The plan of this paper is as follows:\\
In section two
we propose
a new model composed of two different gauge singlet scalar fields with coupling to Higgs boson, they can
have mutual interaction as well.
We impose a
discrete symmetry on only one of them. Hence this field will be a candidate for dark matter.
 The other field will provide us the a strong first order phase transition. And we study the
 parameter
space of the model. In section three we discuss
the finite-temperature potential and explain the origin of the strongly first-order phase
transition at the tree level.
 In section four we discuss
the phenomenological implication of our model.
And finally in section five we present our conclusions.
\section{The Model}
We propose a new extension of the $SM$ by addition of two real gauge
singlet scalars $S_1$ and $S_2$. The Lagrangian of the scalar sector of our model is
given by
\begin{equation}
L_s=(D^\mu H)^\dag(D_\mu H)+\frac{1}{2}\partial^\mu S_1\partial_\mu S_1
+\frac{1}{2}\partial^\mu S_2\partial_\mu S_2
-V(H,S_1,S_2).
\end{equation}
where H denotes the complex Higgs doublet, $H^T=(\chi_1+i\chi_2,\varphi+i\chi_3)/\sqrt{2}.\\$
 As pointed out in Ref. $[11]$,
the most general (renormalizable) tree-level potential for the $SM$
Higgs field $H$ and the singlet $S_1$ depends on $8$ parameters.
 Three more parameters are needed for the field $S_2$, which has a $Z_2$ symmetry.
And another extra parameter as we allow interaction between the gauge singlets
$S_1$ and $S_1$. Hence,
our potential depends on $12$ parameters and it is given by
\begin{eqnarray}
V(H,S_1,S_2)&&=-m^2H^\dag H+\lambda(H^\dag H)^2+\kappa_0S_1+2(\kappa_1S_1+\kappa_2S^2_1+
\kappa_3S^2_2)H^\dag H
\nonumber\\
&&+\frac{1}{2}m^2_1S_1^2+\frac{\lambda_1}{4}S^4_1
+\kappa_4S^3_1+
\frac{1}{2}m^2_2S_2^2+
\frac{\lambda_2}{4}S^4_2+\kappa_5S_1S^2_2
\end{eqnarray}
However, one can always remove the linear term $\kappa_0S_1$ by a redefinition of the
field $S_1$ by a constant shift $[11]$. Therefore, we do not include this linear term
in this work.\\
After symmetry breaking the fields $\chi_1, \chi_2, \chi_2$
becomes the longitudinal degrees of freedom of the weak gauge bosons.
The field $S_1$ is to improve the strength of the phase transition.
And the singlet $S_2$ has a $Z_2$ symmetry. Hence this singlet is a a dark matter candidate.
Therefore at zero temperature the vacuum associated with this field must also be $Z_2$ symmetric.
Hence vacuum expectation value $(vev)$ of this field must vanish.\\
 At $T=0$ we can parameterize the scalar fields of our model by
\begin{equation}
H=\left(%
\begin{array}{c}
  0 \\
  \frac{h+v}{\sqrt{2}}
\end{array}%
\right),\qquad S_1=s_1+x\qquad and \qquad  S_2=s_2,
\end{equation}
with $v\simeq246 (GeV)$ and the parameter $x$ is the $vev$ of the gauge singlet $S_1$.\\
By expanding around the minimum we obtain the squared mass matrix
\begin{equation}
M^2=\left(%
\begin{array}{ccc}
  2\lambda v^2 & 2\kappa_1v +4\kappa_2s_1v& 0\\
  2\kappa_1v +4\kappa_2s_1v& m_1^2+3\lambda_1s_1^2+2\kappa_2v^2+6\kappa_4s_1 & 0 \\
  0 & 0& m_2^2+2\kappa_3 v^2+2\kappa_5s_1\\
\end{array}%
\right).
\end{equation}
At zero temperature the effective potential of the scalar sector of our model is\\
\begin{eqnarray}
V(T=0)&=&-DT^2_0h^2+\frac{\lambda}{4}h^4+\frac{1}{2}m^2_1s_1^2+
\frac{\lambda_1}{4}s_1^4+\kappa_1 h^2 s_1+\kappa_2h^2 s^2_1+\kappa_4 s^3_1
\nonumber\\
&+&\frac{1}{2}m^2_2s_2^2+\kappa_3 h^2 s^2_2+\kappa_5 s_1s_2^2
+\frac{\lambda_2}{4}s_2^4.
\end{eqnarray}
It is convenient to express
the parameters of the potential in terms of the physical masses $M_1$ and $M_2$ of the singlet
scalars, the mixing angle $\theta$ of the singlet $S_1$ and the Higgs field. From the
mass matrix squared we get
\begin{eqnarray}
&&m^2_1=M^2_1 \cos^2(\theta)+M^2_H\sin^2(\theta)-3\lambda_1s_1^2-2\kappa_2v^2-6\kappa_4s_1,
\nonumber\\
&& M_2^2=m_2^2+2\kappa_3v^2+2\kappa_5s_1,
 \qquad
\lambda=\frac{M_H^2\cos^2(\theta)+M^2_1\sin^2(\theta)}{2v^2},
\nonumber\\
&&4(\kappa_1+2\kappa_2s_1)v=(M^2_H-M^2_1)\sin(2\theta)
\end{eqnarray}
and by minimizing the scalar potential we obtain
\begin{eqnarray}
DT^2_0&=&\frac{\lambda v^2+2\kappa_1s_1+2\kappa_2s_1^2}{2}
\nonumber\\
0&=&m_1^2s_1+\lambda_1s_1^3+\kappa_1v^2+2\kappa_2v^2s_1+3\kappa_4s_1^2
\end{eqnarray}
Hence
\begin{equation}
\lambda=\lambda_{SM}-\frac{(M_H^2-M_1^2)\sin^2(\theta)}{2v^2}.
\end{equation}
\subsection{The parameter space of the model}
From previous section we know that the parameter space of the model consists of $(\theta$, $M_1^2$,
$M_2^2$, $\kappa_1$, $\kappa_2$ $\kappa_3$, $\kappa_4$, $\kappa_5$ and $\lambda_2)$.\\
In order to have a stable potential we must have $[11]$
\begin{eqnarray}
&\lambda& >0, \qquad \lambda_1 >0, \qquad \lambda_2 >0,
\nonumber\\
&\kappa_2&>-\frac{\sqrt{\lambda \lambda_1 }}{2}\qquad
and \qquad \kappa_3>-\frac{\sqrt{\lambda \lambda_2 }}{2}.
\end{eqnarray}
But,
if all the eigenvalues of the mass matrix eq. $(4)$ are positive
then the corresponding extremum point is a local minimum.
 Thus, $m_1^2+3\lambda_1s_1^2+2\kappa_2v^2+6\kappa_4s_1>0$.\\
Similarly
 $m_2^2+2\kappa_3 v^2+2\kappa_5s_1>0$.
Hence from eq. $(9)$we have
\begin{equation}
m_1^2+3\lambda_1s_1^2+6\kappa_4s_1>-v^2\sqrt{\lambda \lambda_1}
\qquad and \qquad
 m^2_2+2\kappa_5s_1>-v^2\sqrt{\lambda \lambda_2} .
\end{equation}
From eqs. $(9,10)$ and in terms of the parameters of our model we obtain
\begin{equation}
-\frac{\sqrt{\lambda \lambda_1 }}{2}<\kappa_2<\frac{M^2_1}{2v^2}+\frac{\sqrt{\lambda \lambda_1 }}{2}
\qquad and \qquad
-\frac{\sqrt{\lambda \lambda_2 }}{2}<\kappa_3<\frac{M^2_2}{2v^2}+\frac{\sqrt{\lambda \lambda_2 }}{2}.
\end{equation}
 With $\lambda_{SM}=0.131$, $M_H=126 (GeV)$ and $v=246 (GeV)$
  the ranges of allowed values of the Higgs boson quartic coupling from $eq.(8)$ are
\begin{eqnarray}
\lambda<0.131 \qquad if \qquad M_1<126 (GeV),
\nonumber\\
\lambda=0.131 \qquad if \qquad M_1=126 (GeV),
\nonumber\\
\lambda>0.131 \qquad if \qquad M_1>126 (GeV).
\nonumber\\
\end{eqnarray}
In order to take the effect of mixing angle we notice that,
for a light singlet $(M_1\ll M_H)$ and the mixing angle $\cos (\theta)=0.95$.
From eq. $(8)$ we obtain $\lambda\simeq0.12$ and when $\cos (\theta)=0.99$
$\lambda\simeq0.10$. Here the Higgs boson self coupling is suppressed.\\
For a heavy singlet $(M_1=250 GeV)$ and the mixing angle $\cos (\theta)=0.95$
 we obtain $\lambda\simeq0.25$ and when $\cos (\theta)=0.99$ the value of Higgs boson self coupling
$\lambda\simeq0.19$, Hence the Higgs boson self coupling is enhanced in this region.\\
Therefore, our model predicts variations of the Higgs boson quartic coupling from that of
the $SM$
to be tested by precision
measurements.\\
The triviality bound of the two singlets model is addressed in Ref. $[34]$ and
 quartic coupling remain positive up to energy scalae of about $10 TeV$.\\
In addition to the above theoretical bounds, there are constraints from
experiments. Here we present ranges for the parameters of our model
\begin{eqnarray}
&&-250 GeV <\kappa_1<250 GeV,\qquad-333 GeV <\kappa_4<333 GeV
\nonumber\\
&&-0.25<\kappa_2<0.25,\qquad 0.0001<\kappa_3<0.0025,\qquad
5GeV<M_1<650 GeV,
\nonumber\\
&&0.95<\cos(\theta)<1, \qquad 0<\lambda<0.3,\qquad
0<\lambda_1<4,\qquad 0<\lambda_2<4.
\end{eqnarray}
\section{A strongly first order phase transition and experimental constraints}
At high temperature the effective potential is
\begin{eqnarray}
V(T)&=&D(T^2-T^2_0)h^2-ETh^3
+\frac{\lambda_T}{4}h^4+\frac{1}{2}m^2_1s_1^2+\frac{\lambda_1}{4}s_1^4
+\kappa_1h^2 s_1
\nonumber\\
&+&\kappa_2 h^2 s^2_1+\kappa_4 s^3_1+\frac{1}{2}m^2_2s_2^2
+\frac{\lambda_2}{4}s_2^4+\kappa_3 h^2 s^2_2+
\kappa_5 s_1 s^2_2
\nonumber\\
&+&
[(8\kappa_1+6\kappa_4+8\kappa_5)s_1+(8\kappa_2+3\lambda_1)s^2_1+
(8\kappa_3+3\lambda_2)s^2_2]\frac{T^2}{24},
\end{eqnarray}
The parameters of $eq.(14)$ are given by\\
\begin{eqnarray}
D&=&\frac{1}{24 v^2}(6m^2_W+3m^2_Z+6m^2_t+6\lambda v^2+2(\kappa_2+
\kappa_3)v^2)
\nonumber\\
E&=&\frac{1}{8\pi v^3}(4m^3_W+2m^3_Z)
\nonumber\\
\lambda_T&=&\lambda-\frac{1}{16\pi^2 v^4}(6m^4_Wln\frac{m^2_w}{a_BT^2}
+3m^4_Zln\frac{m^2_Z}{a_BT^2}-12m^4_tln\frac{m^2_t}{a_FT^2})
\nonumber\\
ln a_B&=&3.91, \qquad ln a_F=1.14.
\end{eqnarray}
Let us consider the shape of the potential. As the universe cools down at temperature above
a critical temperature $T_c$ the potential has an absolute minima. At $T_c$ we have two degenerate
minima. The symmetric vacuum is denoted by $(0,s_{1T}, 0)$ and the true vacuum is
 designated by $(h_c,s_{1c}, 0)$, where we
 assumed that the field $S_2$ develops a $vev$ at a temperature above $T_c$ $[13]$.
 The critical temperature is the on-set of the electroweak symmetry
breaking, and a first order $EWPT$ occurs from the symmetric vacuum
to the true vacuum. A transition is considered as strong if $\xi=\frac{h_c}{T_c}>0.6-1.6$,
but this ratio for the $SM$ is $\frac{2E}{\lambda_{SM}}=0.23$.\\
In order to have a symmetric vacua
\begin{equation}
m_1^2s_{1T}+\lambda_1s^3_{1T}+3\kappa_4s^2_{1T}+\frac{1}{12}[(3\lambda_1+8\kappa_2)s_{1T}
+(4\kappa_1+3\kappa_4+4\kappa_5)]T_c^2=0
\end{equation}
The conditions for the existence of the broken vacua are
\begin{equation}
2D(T_c^2-T_0^2)-3ET_ch_c+\lambda_Th^2_c+2\kappa_1s_{1c}+2\kappa_4s_{1c}^2=0,
\end{equation}
and
\begin{eqnarray}
&&m_1^2s_{1c}+\lambda_1s^3_{1c}+3\kappa_4s^2_{1c}+
\kappa_1h_c^2+2\kappa_2s_{1c}h_c^2
\nonumber\\
&&+\frac{1}{12}[(3\lambda_1+8\kappa_2)s_{1c}
+(4\kappa_1+3\kappa_4+4\kappa_5)]T_c^2=0.
\end{eqnarray}
Finally, to have a pair of degenerate vacua the following expression must holds
\begin{eqnarray}
&&\frac{1}{2}m_1^2(s^2_{1T}-s^2_{1c})
+\frac{\lambda_1}{4}(s^4_{1T}-s^4_{1c})
+\kappa_4(s^3_{1T}-s^3_{1c})
\nonumber\\
&&+\frac{1}{24}[(3\lambda_1+8\kappa_2)(s^2_{1T}-s^2_{1c})
+(8\kappa_1+6\kappa_4+8\kappa_5)(s_{1T}-s_{1c})]T_c^2=
\nonumber\\
&&D(T_c^2-T_0^2)h^2_c-ET_ch^3_c+\frac{\lambda_T}{4}h^4_c
+\kappa_1s_{1c}h_c^2+\kappa_2s^2_{1c}h_c^2.
\end{eqnarray}
By solving eqs.(16-19) one can determine the variables $s_{1T}, s_{1c}, T_c$ and $h_c$.\\
The model has a rich parameter space and it is possible to generate a strong first order $EWPT$
with a critical temperature varying from twenty $GeV$ up to few hundred $GeV$.
\section{Phenomenological implications}
In order to investigate the physical implications of the model, we consider two different cases.

\subsection{Gauge singlets without mutual interactions}
In this case $\kappa_5=0$.\\
\underline{\bf {Strong first order EWPT}}\\
In table $1$ we present several configurations. The values of the
parameters $\kappa_3$ and $\cos(\theta)$  for this table are:
\begin{equation}
 \kappa_3=0.001,\qquad\cos(\theta)=0.954.
\end{equation}
Since our numerical results are obtained by using the high temperature-expanded potential,
we explore models which yields critical temperature which are higher than the
mass of the top quark. Hence we expect that high temperature approximation to be
 quantitatively reasonable.\\
 The value of the parameter $\kappa_3$
is chosen from Refs. $[15,35]$, in order to incorporate the most recent experimental
and theoretical constraints. Due to the small value of $\kappa_3$, the dark sector
has little influence on the dynamics which leads to a strong $EWPT$.
We also selected the value of the parameter $\cos(\theta)$
from Ref. $[36]$.\\
The parameter $\beta=\frac{|V_{min}(T=0)|}{|V_{min}(T=T_c)|}$. The
results of table $1$ shows that for lower values of $M_1$, the
electroweak broken vacuum at zero temperature lies much deeper than
the electroweak broken vacuum at the critical temperature.\\
In the last configuration of table $1$, $T_c=196$. Within the range
of validity of high temperature expansion, we compute the
temperature evolution of the $vevs$ of the doublet and the singlet
field up to $T_c$. We also evaluate the temperature evolution of the
electroweak broken vacuum. The results are presented in table $2$.
The parameter $\beta_1=\frac{|V_{min}(T=0)|}{|V_{min}(T=T)|}$ and
for this configuration
$V_{min}(T=0)=-5.568654133\times10^9 (GeV)$.\\

\underline{\bf {Dark matter considerations}}\\
As far as the issue of dark matter is concerned we notice that, the
bounds of the parameter $\kappa_3$ expressed in eq. $(11)$ are from
the stability of the potential and from the positivity of the mass squared matrix.\\
At present a region of interest for the mass of dark matter is the region
$56.8 (GeV)<M_2<63 (GeV)$ $[35]$. We see that, a severe bound of this parameter
in this region can be obtained from the decay width of $H\rightarrow S_2S_2$. In Ref. $[13]$
it is found that
\begin{equation}
\kappa_3\leq 0.013(\frac{GeV}{0.5M_H-M_2})^{\frac{1}{4}}.
\end{equation}
With $M_2=60 (GeV)$ we get $\kappa_3\leq 0.0096$.\\
Yet a more restrictive bound on this parameter (see Ref. $[35]$) based on the relic density of dark
matter, which comes from experiments is
$0.0000625<\kappa_3<0.00125$.\\
While in Refs. $[13,15]$ these results, which prevent
 the occurrence of a strong first order $(EWPT)$ is catastrophic, in our
model the small values of $\kappa_3$ are acceptable. In fact from $eq.(15)$
we see that smaller values of $\kappa_3$ are preferred as in this case the scalar $S_2$
will have a minor role in the occurrence of a strong first order $EWPT$.\\
Moreover, we see that in this region the variation
of the parameter $\lambda$ due to singlet scalar $S_1$ does not have influence
on the bounds for $\kappa_3$.\\
Hence as far as dark matter is concerned,
we find that the situation is very similar to that of Ref. $[15]$.\\
\subsection{Gauge singlets with mutual interactions}
In this case the gauge singlets are allowed to interact with each other directly. \
By considering the annihilation of $DM$ into $SM$ particles one must include
the effect of a s channel reaction mediated by the gauge singlet $S_1$,
which has an amplitude proportional to $\frac{\kappa_1\kappa_5}{s-M^2_1+iM_1\Gamma_{s1}}$.\\
The parameter $\kappa_5$ must chosen in a way that the extra $s$
channel reaction does not alter the dark matter cross section
significantly, hence
  we defer an investigation of this matter to a future work.\\
\section{Conclusions}
There have been some attempts to address the issues of dark matter and
occurrence of a strong first order electroweak phase transition in a single unified model $[13]$.
However, we find that it is difficult to solve both problems with a single scalar.
Hence, in this paper we have presented a new two singlet scalar model.\\
In our model, when
  the dark matter
field $S_2$ is only coupled to Higgs boson, we find that the phenomenology of the dark sector of the
model is similar to that of a singlet scalar dark matter.\\
We have explored the characteristics of a strong first order $EWPT$
in the mass range $150 GeV<M_1<550 GeV$.
\\\\
{\bf Acknowledgment}:\\
We are very grateful to the referee for inspiring suggestions and comments.\\

\clearpage
Table Captions:\\\\

Table $1$ : The values of the critical temperature and the strength
of a strong first order $EWPT$ for several configurations. The
values of $\kappa_3$ and  $cos(\theta)$are within the current
experimental bounds.\\\\
Table $2$ : Temperature evolution of the $vevs$ of the doublet $(h)$
and the singlet field $(s_1)$ up to $T_c$, for the case of $M_1=550
(GeV)$.
 The parameter
$\beta_1$ is a measure of the location of the electroweak broken
vacuum at
temperature $T$.\\\\

\clearpage

Table $1$\\\\\

\begin{tabular}{|c|c|c|c|c|c|c|}
  \hline
  $M_1(GeV)$ & $\kappa_1$ & $\kappa_2$ & $\kappa_4$ & $T_c(GeV)$ & $\beta$ & $\xi$ \\
  \hline
  150 & -14.30 & -0.0510 & -0.107 & 174.0 & 8.65 & 1.17 \\
  250 & -14.30 & 0.0625 & -0.089 & 177.9 & 5.67 & 1.12 \\
  350 & -10.70 & 0.2000 & 0.080 & 183.9 & 3.14 & 1.08 \\
  450 & -7.20 & 0.2490 & 0.212 & 190.8 & 1.71 & 1.04 \\
  550 & -5.23 & 0.2499 & 0.237 & 196.0 & 1.30 & 1.02 \\
  \hline
\end{tabular}

\clearpage

Table $2$\\\\\

\begin{tabular}{|c|c|c|c|}
  \hline
  T (GeV)& $s_1$ & h & $\beta_1$ \\
  \hline
  174.0 & 308.34 & 242.10 & 1.22\\
  179.5 & 309.68 & 233.09 & 1.24 \\
  185.0 & 311.14 & 223.08 & 1.26 \\
  190.5 & 312.72 & 211.95 & 1.28 \\
  196.0 & 314.43 & 199.46 & 1.30 \\
  \hline
\end{tabular}


\clearpage
\begin{thebibliography}{14}
\bibitem{1}
A. Sakharov, Pisma Zh. Eksp. Teor. Fiz. {\bf 5} (1967) 32.
\bibitem{2}
V. A. Kuzmin, V. A. Rubakov and M. E. Shaposhnikov, Phys. Lett. B {\bf 155} (1985) 36.
\bibitem{3}
G. W. Anderson and L. J. Hall, Phys. Rev. D {\bf 45} (1992) 2685.
\bibitem{4}
K. G. Chetyrkin and M. F. Zoller, JHEP {\bf 1304} (2013) 091.
\bibitem{5}
J. McDonald, Phys. Rev. D {\bf 50 } (1994) 3637.
\bibitem{6}
J. J. van der Bij, Phys. Lett. B {\bf 636} (2006) 56.
\bibitem{7}
S. Profumo, M. J. Ramsey-Musolf, G. Shaughnessy, JHEP {\bf 0708} (2007) 010.
\bibitem{8}
A. Noble and M. Perelstein, Phys. Rev. D {\bf 78} (2008) 063518.
\bibitem{9}
V. Barger, P. Langacker, M. McCaskey, M. J. Ramsy-Musolf and G. Shaughnessy,
Phys. Rev. D {\bf 77} (2008) 035005.
\bibitem{10}
S. Das, P. J. Fox, A. Kumar and N. Weiner, JHEP {\bf 1011} (2010) 108.
\bibitem{11}
J. R. Espinosa, T. Konstandin, and F. Riva, Nucl. Phys. B {\bf 854} (2012) 592.
\bibitem{12}
J. M. Cline, K. Kainulainen, P. Scott and C. Weniger, Phys. Rev. D {\bf 88}  (2013) 055025.
\bibitem{13}
J. M. Cline, K. Kainulainen, JCAP {\bf 1301} (2013) 012.
\bibitem{14}
L. Basso, O. Fisher and J. J. van der Bij, Phys. Lett. B {\bf 730} (2014) 326.
\bibitem{15}
T. Alanne, K. Tuominen, and V. Vaskonen, Nucl. Phys. B {\bf 889} (2014) 692.
\bibitem{16}
A. Katz and M. Perelstein, JHEP {\bf 1407} (2014) 108 .
\bibitem{17}
D. J. Chung, A. J. Long, and L.-T. Wang, Phys. Rev. D {\bf 87} (2013) 023509.
\bibitem{18}
 C. Caprini and R. Durrer, Phys. Rev. D {\bf74}  (2006) 063521.
\bibitem{19}
 A. Kosowsky, A. Mack and T. Kahniashvili, Phys. Rev. D {\bf66} (2002) 024030.
\bibitem{20}
 A. Kosowsky, M. S. Turner and R. Watkins, Phys. Rev. D {\bf 45} (1992) 4514.
\bibitem{21}
 M. Kamionkowski, A. Kosowsky and M. S. Turner, Phys. Rev. D {\bf 49} (1994) 2837.
\bibitem{22}
J. R. Espinosa and M. Quiros, Phys. Rev. D {\bf 76} (2007) 07600.
\bibitem{23}
J. R. Espinosa, T. Konstandin, J. M. No and M. Quiros, Phys. Rev. D {\bf 78}(2008) 123528.
\bibitem{24}
A. Drozd, B. Grzadkowski and J. Wudka, Acta Phys. Polon. B {\bf 42} (2011) 2255.
\bibitem{25}
A. Abada, S. Nasri and D. Ghaffor, Phys. Rev. D {\bf 83} (2011) 095021.
\bibitem{26}
A. Drozd, B. Grzadkowski and J. Wudka, JHEP {\bf 1204} (2012) 006.
\bibitem{27}
A. Ahriche, A. Arhrib and S. Nasri, JHEP {\bf 02} (2014) 042.
\bibitem{28}
D. Curtin, P. Meade and C. T. Yu, JHEP {\bf 1411} (2014) 127.
\bibitem{29}
G. M. Pruna and T. Robens, Phys. Rev. D {\bf 88} (2013)115012.
\bibitem{30}
L. Dolan and R. Jackiw, Phys. Rev. D {\bf 9} (1974) 3320.
\bibitem{31}
E. W. Kolb and M. S. Turner, \textit {The Early Universe}, Front. Phys. {\bf 69} (1990) 1–547.
\bibitem{32}
J. I. Kapusta, \textit{Finite temperature field theory}, Cambridge University Press, (1989).
\bibitem{33}
M. Le-Bellac, \textit{Thermal field theory}, Cambridge University Press, (2000).
\bibitem{34}
A. Abadaa and S. Nasri, Phys. Rev. D {\bf 88} (2013) 016006.
\bibitem{35}
L. Feng, A. S. Profumo and B. L. Ubaldic, arXiv:hep-ph/1412.1105v1 (2014).
\bibitem{36}
S. Profumo, M J. Ramsey-Musolf, C. L. Wainwright and P. Winslow, arXiv:hep-ph/1407.5342v1 (2014).






\end{thebibliography}
\end{document}